\documentclass[graybox, fleqn]{svmult}

\usepackage[utf8]{inputenc}

\usepackage{mathptmx}       
\usepackage{helvet}         
\usepackage{courier}        
\usepackage{type1cm}        
\usepackage{makeidx}         
\usepackage{graphicx}        
\usepackage{multicol}        
\usepackage[bottom]{footmisc}

\usepackage[numbers]{natbib}

\usepackage{amsmath}
\usepackage{amssymb}
\usepackage{todonotes}
\usepackage{tikz}
\usepackage{caption}
\usepackage{tabularx}
\usepackage{booktabs}

\usetikzlibrary{shapes,arrows,positioning}

\makeindex             

\begin{document}

\title*{Analysis of Temporal Features for Interaction Quality Estimation}

\titlerunning{Analysis of Temporal Features for IQ Estimation}

%

\author{Stefan Ultes, Alexander Schmitt, and Wolfgang Minker}
\authorrunning{S. Ultes, A. Schmitt, and W. Minker}
\institute{Stefan Ultes \at Cambridge University Engineering Department, UK
 \and
 Alexander Schmitt \at Ulm University, Ulm, Germany
 \and
 Wolfgang Minker \at Ulm University, Ulm, Germany}
\maketitle

\abstract{Many different approaches for estimating the Interaction Quality (IQ) of Spoken Dialogue Systems have been investigated. While dialogues clearly have a sequential nature, statistical classification approaches designed for sequential problems do not seem to work better on automatic IQ estimation than static approaches, i.e., regarding each turn as being independent of the corresponding dialogue. Hence, we analyse this effect by investigating the subset of temporal features used as input for statistical classification of IQ. We extend the set of temporal features to contain the system and the user view. We determine the contribution of each feature sub-group showing that temporal features contribute most to the classification performance. Furthermore, for the feature sub-group modeling the temporal effects with a window, we modify the window size increasing the overall performance significantly by +15.69\% achieving an Unweighted Average Recall of 0.562.}

\abstract*{Many different approaches for estimating the Interaction Quality (IQ) of Spoken Dialogue Systems have been investigated. While dialogues clearly have a sequential nature, statistical classification approaches designed for sequential problems do not seem to work better on automatic IQ estimation than static approaches, i.e., regarding each turn as being independent of the corresponding dialogue. Hence, we analyse this effect by investigating the subset of temporal features used as input for statistical classification of IQ. We extend the set of temporal features to contain the system and the user view. We determine the contribution of each feature sub-group showing that temporal features contribute most to the classification performance. Furthermore, for the feature sub-group modeling the temporal effects with a window, we modify the window size increasing the overall performance significantly by +15.69\% achieving an Unweighted Average Recall of 0.562.}

\section{Introduction}
%

Due to recent advances in Speech Recognition Technology, technical systems with speech interfaces are becoming more and more prevalent in our everyday lives. 
Comparing the performance of such SDSs is a non-trivial task which has not yet been solved. While many (mostly statistical) approaches on spoken dialogue assessment take into account objective criteria like dialogue length or task success rate, the focus has shifted to more user-centered criteria, such as the satisfaction level measured while or after users have interacted with an SDS.

To achieve automatic evaluation, recent approaches focus on User Satisfaction (US) recognition employing state-of-the-art statistical classification systems. While some of these approaches deal with US on the dialogue level, i.e., providing a satisfaction score for the complete interaction, recent work focuses on US on the system-user-exchange level. Having this information, the system may react adaptively depending on the current US score~\cite{ultes2011b,ultes2014,ultes2014a,ultes2015}.

In this work, we focus on a US-based exchange-level measure called Interaction Quality (IQ)~\cite{schmitt2015}. Naturally, having a dialogue consisting of several turns (or system-user-exchanges) in a temporal order may evidently be regarded as a sequence of exchanges. Furthermore, the IQ score on the exchange-level measuring the quality up to a single interaction step highly depends on the IQ scores of the previous dialogue steps~\cite{ultes2014b}. For such a sequential problem, many classification approaches specifically designed for the needs inherent in sequential problems exist, e.g., Hidden Markov Models (HMMs)~\cite{rabiner1989}. However, applying these approaches to IQ estimation has not resulted in an increase in performance compared static approaches\footnote{Regarding each exchange being independent of all other exchanges, and not as part of a sequence.}, e.g., applying a support vector machine (SVM)~\cite{vapnik1995}. First experiments applying a Conditioned Random Field~\cite{lafferty2001} or Recurrent Neural Networks~\cite{hochreiter1997}, which have both shown to achieve good performance for other tasks, also resulted in low performance.


Thus, while the problem of estimating the Interaction Quality clearly seems as if it would benefit from sequential classification, it is not a straight forward problem to outperform static approaches. Here, we assume that the reason why static approaches perform that well lies in the modelling of the interaction parameters.
They consist of a high number of temporal features thus encoding temporal information about the dialogue.
However, simply adding the previous IQ value to the feature set is not sufficient and does result in worse performance~\cite{ultes2014b}. To prove our assumption, these temporal features are analyzed.  

Additionally, the set of temporal features is extended to contain both the system and the user view and some temporal features are modified. Thus, in contrast to previous work where different classification approaches have been investigated, we aim at increasing the estimation performance by extending and optimizing the feature set.

The remainder of this paper is organized as follows: significant related work on user satisfaction recognition in general as well as on Interaction Quality estimation specifically is stated in Section~\ref{sec:relatedWork}. Our main contribution of analyzing the temporal features for IQ estimation is presented in Section~\ref{sec:evaluation} including a thorough description of the Interaction Quality paradigm. Furthermore, in the same section, we will argue for a recalculation of some features and describe the new extended feature set. We compare the newly created feature set with the original features for different feature sub-groups as well as analyse the performance analysis of different temporal contexts. The results are discussed in Section~\ref{sec:discussion}. Finally, the results are interpreted in Section~\ref{sec:conclusion} and future work is outlined.



\section{Significant Related Work}
\label{sec:relatedWork}

Estimating User Satisfaction (US) for SDSs has been in the focus of research for many years. In this section, we present approaches on US recognition in general and on IQ estimation specifically.

Groundbreaking work on automatic SDS evaluation has been presented by Walker et al.~\cite{walker1997} with the PARADISE framework. The authors assume a linear dependency between quantitative parameters derived from the dialogue and User Satisfaction on the dialogue level, modeling this dependency using linear regression. Unfortunately, for generating the regression model, weighting factors have to be computed for each system anew. This generates high costs as dialogues have to be performed with real users where each user further has to complete a questionnaire after completing the dialogue. Moreover, in the PARADISE framework, only quality measurement for whole dialogues (or system) is allowed. However, this is not suitable for online adaptation of the dialogue~\cite{ultes2012}. Furthermore, PARADISE relies on questionnaires while we focus on work using single-valued ratings.

Numerous work on predicting User Satisfaction as a single-valued rating task for each system-user-exchange has been performed using both static and sequential approaches. 
Hara et al.~\cite{hara2010} derived turn level ratings from an overall score applied by the users after the dialogue. Using n-gram models reflecting the dialogue history, the estimation results for US on a 5 point scale showed to be hardly above chance.

Higashinaka et al.~\cite{higashinaka2010b} proposed a model to predict turn-wise ratings for human-human dialogues (transcribed conversation) and human-machine dialogues (text from chat system). Ratings ranging from 1-7 were applied by two expert raters labeling ``Smoothness'', ``Closeness'', and ``Willingness'' not achieving a Match Rate per Rating (MR/R)
of more than 0.2-0.24 applying Hidden Markov Modes as well as Conditioned Random Fields. These results are only slightly above the random baseline of 0.14. Further work by Higashinaka et al.~\cite{higashinaka2010a} uses ratings for overall dialogues to predict ratings for each system-user-exchange using HMMs. Again, evaluating in three user satisfaction categories ``Smoothness'', ``Closeness'', and ``Willingness'' with ratings ranging from 1-7 achieved best performance of 0.19 MR/R.

An approach presented by Engelbrecht et al.~\cite{engelbrecht2009} uses Hidden Markov Models (HMMs) to model the SDS as a process evolving over time. User Satisfaction was predicted at any point within the dialogue on a 5 point scale. Evaluation was performed based on labels the users applied themselves during the dialogue.

Work by Schmitt et al.~\cite{schmitt2011e} deals with determining User Satisfaction from ratings applied by the users themselves during the dialogues. A Support Vector Machine (SVM) was trained using automatically derived interaction parameter to predict User Satisfaction for each system-user-exchange on a 5-point scale achieving an MR/R 
of 0.49.

Interaction Quality (IQ) has been proposed by Schmitt et al.~\cite{schmitt2015} as an alternative performance measure to US. In their terminology, US ratings are only applied by users. As their presented measure uses ratings applied by expert raters, a different term is used. Using the same approach as for estimating US, they achieve an MR/R
of 0.58 for estimating IQ on the turn level on a 5-point scale. The IQ paradigm will be described in more detail in Section~\ref{sec:interactionQuality}.

To improve the performance of static classifiers for IQ recognition, Ultes et al.~\cite{ultes2013d} proposed a hierarchical approach: first, IQ is predicted using a static classifier. Then, the prediction error is calculated and a second classifier is trained targeting the error value. In a final step, the initial hypothesis may then be corrected by the estimated error. This approach has been successfully applied improving the recognition performance relatively by up to +4.1\% Unweighted Average Recall (UAR, the average of all class-wise recalls).

Work on rendering IQ prediction as a sequential task analyzing HMMs and Conditioned Hidden Markov Models has been performed by Ultes et al.~\cite{ultes2012b}. They achieved an UAR\ of 0.39 for CHMMs. This was outperformed by regular HMMs (0.44 UAR) using Gaussian mixture models for modeling the observation probability for both approaches. Replacing the observation probability model with the confidence scores of static classification methods, Ultes et al.~\cite{ultes2014b} achieved a significant improvement of the baseline with an UAR of 0.51.

Unfortunately, applying classification approaches which render the task of IQ prediction as a sequential problem do not seem to increase the estimation performance. Therefore, as the feature set used for classification also models the temporal effects inherent in IQ estimation to a certain degree, it will be analysed more closely in the following section.

\section{Temporal Feature Analysis}
\label{sec:evaluation}

The main goal of this paper is to analyze why the set of interaction parameters used for IQ estimation is so powerful that most static approaches on IQ estimation outperform sequential ones. Moreover, in this paper, the emphasis lies on the temporal features, i.e., parameters on the window and dialogue levels. First, the Interaction Quality Paradigm including the definition of IQ and a description of the Interaction Parameters is presented followed by a brief description of the used evaluation methods. Then, the temporal subset of those parameters is manually analyzed to find peculiarities. Furthermore, the performance of each level separately as well as the absence of each level is analyzed. Finally, the effect of the window-size on the estimation performance in investigated.

\subsection{The Interaction Quality Paradigm}
\label{sec:interactionQuality}

The Interaction Quality paradigm has been originally introduced by Schmitt et al.~\cite{schmitt2011a}. It represents a scheme for bridging the gap between the subjective nature of user satisfaction and objective ratings. The general idea is to utilize statistical models to predict the Interaction Quality (IQ) of system-user-exchanges based on interaction parameters which are both described in the following. The annotated exchanges along with the interaction parameters have been combined within the LEGO corpus~\cite{schmitt2012a} forming the base of this work.



The Interaction Quality as a measure for dialogue performance intends to overcome the problems inherit with the purely subjective measure user satisfaction. Hence, it is defined similarly to user satisfaction: while the latter represents the true disposition of the user, IQ is the disposition of the user assumed by expert annotators\footnote{IQ is strongly related to user satisfaction~\cite{ultes2013a} achieving a Spearman's $\rho$ of $0.66$ $(\alpha < 0.01)$.}. Here, expert annotators listen to recorded dialogues after the interactions and rate them by assuming the point of view of the actual person performing the dialogue. These experts are supposed to have some experience with dialogue systems. In this work, expert annotators were ``advanced students of computer science and engineering'' \cite{schmitt2011a}, i.e., grad students.

The employed corpus (``LEGO corpus'') is based on 200 calls to the ``Let's Go Bus Information System'' of the Carnegie Mellon University in Pittsburgh~\cite{raux2006} recorded in 2006.
Labels for IQ have been assigned by three expert annotators, achieving a total of 4,885 system-user-exchanges\footnote{A system turn followed by a user turn.} with an inter-annotator agreement of $\kappa = 0.54$. This may be considered as a moderate agreement (cf.~Landis and Koch's Kappa Benchmark Scale~\cite{landis1977}) which is quite good considering the difficulty of the task that required to rate each exchange.
The final label was assigned to each exchange by using the median of all three individual ratings.

The experts applied ratings at a scale from 1 (extremely unsatisfied) to 5 (satisfied) considering the complete dialogue up to the current exchange. Thus, each exchange has been rated without regarding any upcoming user utterance. Each dialogue starts with a rating of 5 since the user is expected to be satisfied in the beginning. To compensate the subjective nature and to ensure consistent labeling, the expert annotators had to follow labeling guidelines \cite{schmitt2012a}.

\begin{figure*}[tb]
 \centering
 \includegraphics[width=\textwidth]{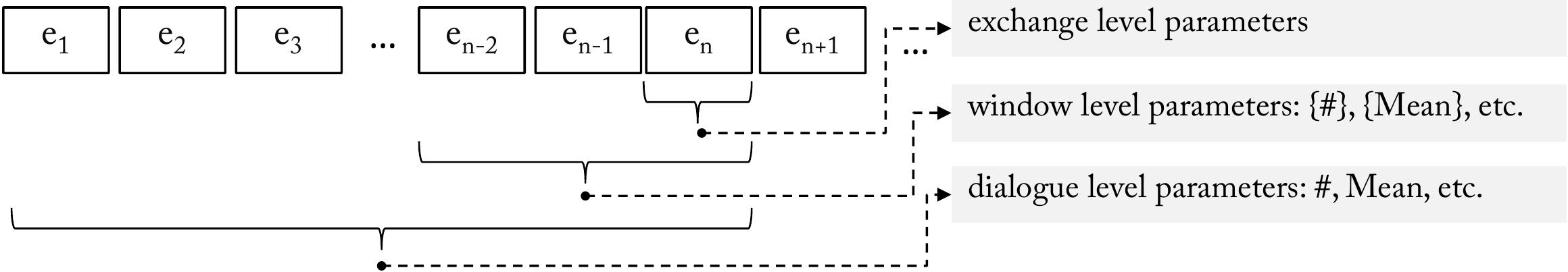}

\caption{The three different modeling levels representing the interaction at
exchange $e_n$: The most detailed exchange level, comprising parameters of the
current exchange; the window level, capturing important parameters from the
previous $n$ dialog steps (here $n=3$); the dialog level, measuring overall
performance values from the entire previous interaction. The figure is taken from Schmitt et al.~\cite{schmitt2011a}.}
\label{fig:modeling_scope}
\end{figure*}

The set of \emph{interaction parameters} used as input variables for the IQ model consists of a total of 53 parameters automatically derived from three SDS modules: Automatic Speech Recognition, Spoken Language Understanding, and Dialogue Management. Furthermore, to account for the temporal nature of the system, the parameters are modeled on three different levels (Figure \ref{fig:modeling_scope}): \textit{Exchange level} parameters are derived directly from the respective dialogue modules, e.g., \texttt{ASRConfidence}. \textit{Dialogue level} parameters consist of counts (\#), means (Mean), etc.\ of the exchange level parameters calculated from all exchanges of the whole dialogue up to the current exchange, e.g., \texttt{MeanASRConfidence}. \textit{Window level} parameters consist of counts (\{\#\}), means (\{Mean\}), etc.\ of the exchange level parameters calculated from the last three exchanges, e.g., \texttt{\{Mean\}ASRConfidence}.
A thorough description of all interaction parameters on all levels can be found in Schmitt et al.~\cite{schmitt2012a}.

\begin{figure*}[t]
 \centering
 \footnotesize
 \begin{tabular}{lccccccc}
 \toprule
 	& & & & \multicolumn{2}{c}{\%ASR-Success} & \multicolumn{2}{c}{\{\#\}ASR-Success} \\
     & System DA & User DA  & ASR-Status & User & System & User & System \\
    \midrule
    1 & [Welcome] & - & - & 0.0 & 0.0 & 0 & 0 \\
    2     &[Help\_info]       &  -                    &   -        & 0.0 & 0.00  & 0 & 0 \\
    3     &[Open]             & [Inform\_origin]& complete   & 1.0 & 0.33 & 1 & 1 \\
    4     &[Confirm\_origin]  &[Inform\_origin]& complete   & 1.0 & 0.50  & 2 & 2 \\
    5     &[Confirm\_origin]  & [Affirm]& incomplete & 0.66 & 0.40  & 2 & 2 \\
    6     &[Filler]           &  -                    &    -       & 0.66 & 0.33  & 2 & 1 \\
    7     &[Ask\_destination] & [Inform\_destination]& complete   & 0.75 & 0.43 & 2 & 1 \\
    \bottomrule
    \end{tabular}%
\caption{The computation of dialog level parameters \%ASR-Success (percentage of successfull ASR events)
and window-level parameters \{\#\}ASR-Success (number of successful ASR events within the window frame) from the view of the user and the system.
}
\label{fig:parameterUserSystem}
\end{figure*}

For measuring the performance of the classification algorithms, we rely on \textit{Unweighted Average Recall (UAR)} as the average of all class-wise recalls,
\textit{Cohen's Kappa}~\cite{cohen1960} linearly weighted~\cite{cohen1968} and \textit{Spearman's Rho}~\cite{spearman04}. The latter two also represent a measure for similarity of paired data.

\subsection{Manual Analysis and Feature Set Extension}

The temporal effects of the dialogue are captured within the interaction parameters by the dialogue and window levels. To get a better understanding, the corpus is analysed more closely. The observation been made is that some system-user-exchanges contain only a system utterance without user input, e.g., the ``Welcome'' message of the system (Figure~\ref{fig:parameterUserSystem}, line 1). While the system and the user have a different view on the interaction in general, this is especially the case regarding the number of dialogue turns. However, as this information is used for computing parameters on the dialogue and window level, both views should be reflected by the interaction parameters. Hence, the parameters should be computed with respect to the number of system turns and with respect to the number of user turns as well. Thus, the original feature set is extended to contain both variants of parameters. 

An example dialogue snippet showing parameters originating from the \emph{ASR-Status} is illustrated in Figure~\ref{fig:parameterUserSystem}. It shows both calculation variants for the window parameter \emph{\{\#\}ASR-Success} and the dialogue parameter \emph{\%ASR-Success}. The differences are clearly visible: while \emph{\%ASR-Success} is either 0 or 1 for the user's view (only successful ASR events occur), the numbers are different for the system's view.

To reflect this system and user view for the complete corpus, a number of parameters are recalculated for both variants\footnote{Recalculated parameters: \%ASRSuccess, \%TimeOutPrompts, \%ASRRejections, \%TimeOuts\_ASRRej, \%Barge-Ins, MeanASRConfidence, \{\#\}ASRSuccess, \{\#\}TimeOutPrompts, \{\#\}ASRRejections, \{\#\}TimeOuts\_ASRRej, \{\#\}Barge-Ins, \{Mean\}ASRConfidence}. The window size remained the same with $n=3$. This results in an extended feature set consisting of 65 features. For the remainder of the paper, we will refer to the original feature set as $LEGO_{orig}$ and to the extended feature set as $LEGO_{ext}$.

\subsection{Analysis of Parameter Levels}
\begin{figure}[t]
 \centering
 \includegraphics[width=0.7\columnwidth]{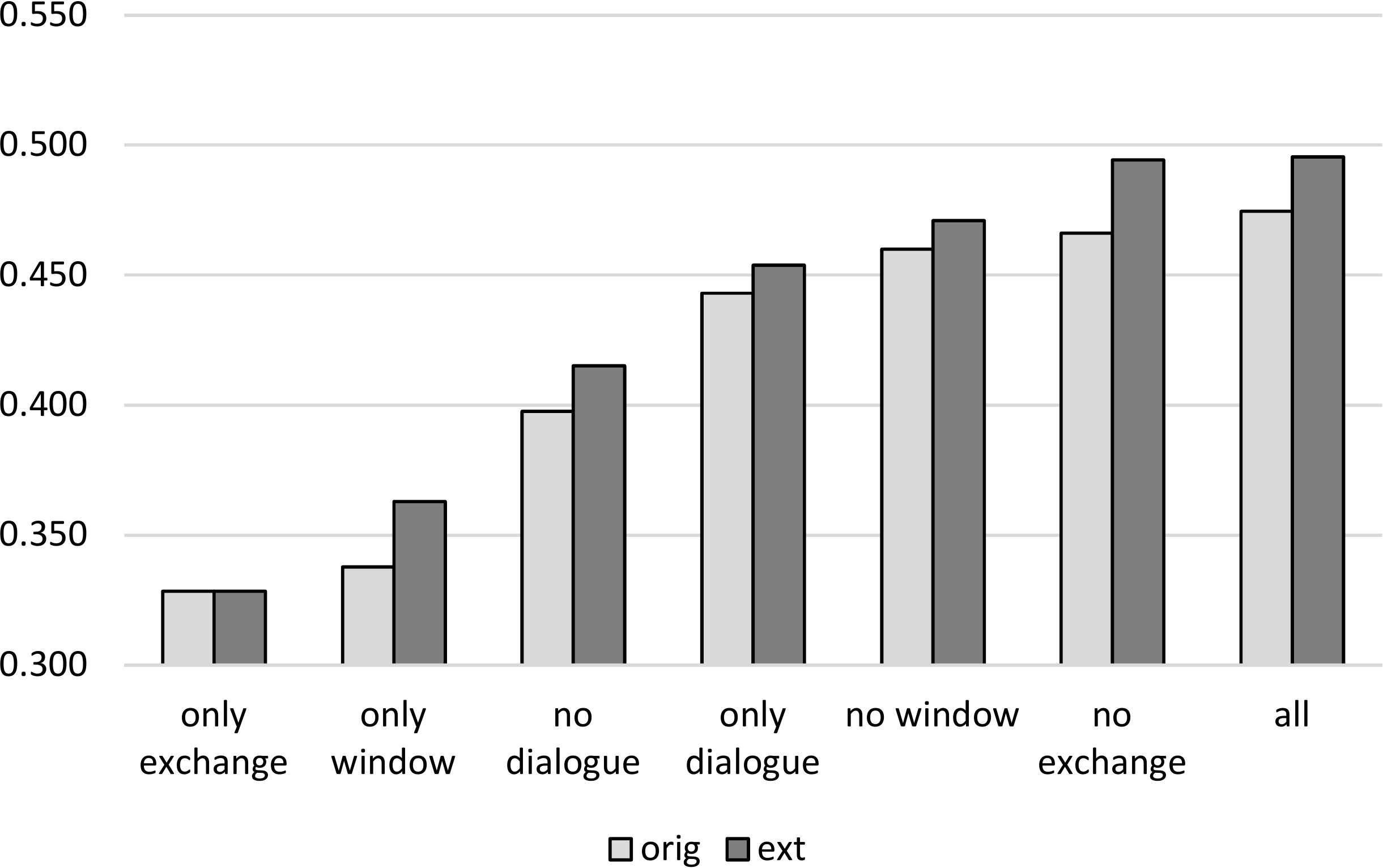}
\caption{SVM performance in UAR for including and excluding different parameter levels for $LEGO_{orig}$ and for $LEGO_{ext}$.}
\label{fig:resultsLevels}
\end{figure}

To get a better understanding of the different parameter level and their contribution to the overall estimation performance, experiments have been conducted using each combination of parameter levels as a feature set, e.g., using only parameters on one level or using parameters from all but one levels. Furthermore, to get a better understanding of the extension of the feature set, the experiments are performed for $LEGO_{orig}$ and for $LEGO_{ext}$. Some interaction parameters with constant value and textual interaction parameters with a task-dependent nature have been discarded\footnote{Discarded parameters: Activity, LoopName, Prompt, SemanticParse, SystemDialogueAct, UserDialogueAct, Utterance, parameters related to modality and help requests on all levels} leaving 38 parameters for $LEGO_{orig}$ and 50 for $LEGO_{ext}$. The results are computed using a linear Support Vector Machine (SVM)~\cite{vapnik1995} in a 10-fold cross-validation setting. The results are stated in Table~\ref{tab:resultsLevels} and visualized in Figure~\ref{fig:resultsLevels}.

\begin{table*}[tb]
  \centering
  \footnotesize
    \begin{tabular}{rcccccc}
\toprule
          & \multicolumn{2}{c}{UAR} & \multicolumn{2}{c}{$\kappa$} & \multicolumn{2}{c}{$\rho$} \\
          \cmidrule(l{2pt}r{2pt}){2-3} \cmidrule(l{2pt}r{2pt}){4-5} \cmidrule(l{2pt}r{2pt}){6-7}
          & $LEGO_{orig}$ & $LEGO_{ext}$ & $LEGO_{orig}$ & $LEGO_{ext}$ & $LEGO_{orig}$ & $LEGO_{ext}$ \\
          
          \midrule
    only exchange & 0.328 & 0.328 & 0.310 & 0.310 & 0.456 & 0.456 \\
    only window & 0.338 & 0.363 & 0.333 & 0.380 & 0.479 & 0.558 \\
    no dialogue & 0.398 & 0.415 & 0.457 & 0.480 & 0.622 & 0.643 \\
    only dialogue & 0.443 & 0.454 & 0.559 & 0.571 & 0.726 & 0.738 \\
    no window & 0.460 & 0.471 & 0.578 & 0.589 & 0.737 & 0.747 \\
    no exchange & 0.466 & 0.494 & 0.584 & 0.611 & 0.747 & 0.764 \\
    all   & 0.475 & 0.495 & 0.596 & 0.616 & 0.757 & 0.770 \\
	\bottomrule

    \end{tabular}%
    \caption{Results in UAR, $\kappa$ and $\rho$ for including and excluding different parameter levels for $LEGO_{orig}$ and for $LEGO_{ext}$.}
  \label{tab:resultsLevels}%
\end{table*}%

Best performance for both $LEGO_{orig}$ and $LEGO_{ext}$ in terms of UAR, $\kappa$, and $\rho$ is achieved by using all parameters. However, it is highly notable that the results are very similar compared to the results of using all but the exchange level parameters (\emph{no exchange}). In fact, applying the Wilcoxon test~\cite{wilcoxon1945} for statistical significance proves the difference to be non-significant ($LEGO_{orig}$: $p>.15$, $LEGO_{ext}$: $p>.94$). This is underpinned by the results of only using the parameters on the exchange level (\emph{only exchange}) being among the worst performing configurations together with the \emph{no window} results. However, comparing the \emph{all} results to the \emph{no window} results ($LEGO_{orig}$: $p<.1$, $LEGO_{ext}$: $p<.001$) reveals that the window parameters play a bigger role in the overall performance.

While the analysis above is true for both feature sets $LEGO_{orig}$ and $LEGO_{ext}$, the results clearly show that the extension of the feature set results in an increased performance on almost all levels. The overall performance using \emph{all} parameters has been relatively increased by 4.4\% ($p<.001$) in UAR and the performance of \emph{no exchange} has been relatively increased by 6.0\% ($p<.001$). The results of the \emph{only exchange} parameters are the same for both feature sets as the parameters on this level are the same, i.e., have not been computed anew.

\subsection{Analysis of Window Size}

\begin{figure*}[t]
 \centering
 \includegraphics[width=.9\textwidth]{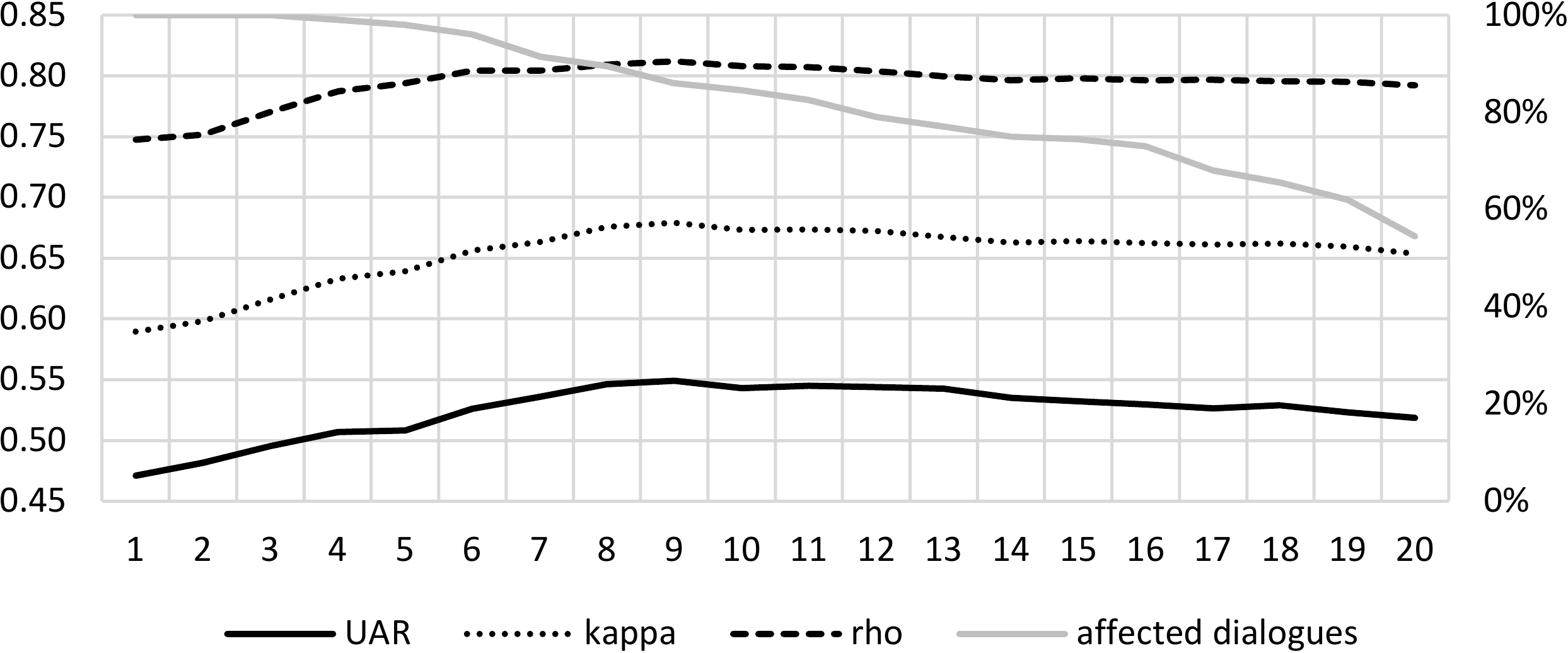}
\caption{SVM performance (left ordinate) for $LEGO_{ext}$ using different window sizes from $n=1$ (no window) to $n=20$ (abscissa). The percentage of affected dialogue, i.e., which have a length greater than the window size, is shown on the right ordinate.}
\label{fig:resultsWindow}
\end{figure*}

\begin{table}[t]
\footnotesize
  \centering
  \setlength{\tabcolsep}{6pt}
    \begin{tabular}{rrlccc}
    \toprule
    Window & \multicolumn{2}{c}{UAR}  & $\kappa$ & $\rho$   & \#dial. \\
    \midrule
    1     & 0.471 & - 4.93\%** & 0.589 & 0.747 & 100\% \\
    2     & 0.482 & - 2.76\%** & 0.598 & 0.752 & 100\% \\
    3     & 0.495 &  \multicolumn{1}{c}{--} & 0.616 & 0.770 & 100\% \\
    4     & 0.507 & + 2.30\% & 0.633 & 0.787 & 99\% \\
    5     & 0.508 & + 2.57\%** & 0.639 & 0.794 & 98\% \\
    6     & 0.526 & + 6.16\%** & 0.656 & 0.804 & 96\% \\
    7     & 0.536 & + 8.16\%** & 0.663 & 0.804 & 92\% \\
    8     & 0.546 & + 10.22\%** & 0.675 & 0.809 & 90\% \\
    9     & 0.549 & + 10.82\%** & 0.679 & 0.812 & 86\% \\
    10    & 0.543 & + 9.61\%** & 0.673 & 0.808 & 85\% \\
    11    & 0.545 & + 9.97\%** & 0.674 & 0.807 & 83\% \\
    12    & 0.544 & + 9.76\%** & 0.672 & 0.804 & 79\% \\
    13    & 0.542 & + 9.50\%** & 0.668 & 0.800 & 77\% \\
    14    & 0.535 & + 7.99\%* & 0.663 & 0.797 & 75\% \\
    15    & 0.532 & + 7.42\%** & 0.664 & 0.798 & 75\% \\
    16    & 0.530 & + 6.90\%** & 0.663 & 0.796 & 73\% \\
    17    & 0.526 & + 6.23\%** & 0.661 & 0.797 & 68\% \\
    18    & 0.529 & + 6.75\%* & 0.662 & 0.796 & 66\% \\
    19    & 0.523 & + 5.54\%* & 0.659 & 0.795 & 62\% \\
    20    & 0.519 & + 4.66\%* & 0.654 & 0.792 & 55\% \\
    \bottomrule
    \end{tabular}%
    \caption{Results of different window sizes for IQ recognition in UAR, $\kappa$, and $\rho$. In addition, the relative improvement in UAR with respect to a window size of 3 is depicted. Significance is indicated with * ($\alpha < 0.05$) and ** ($\alpha < 0.01$) determined using the Wilcoxon test~\cite{wilcoxon1945}. Best performance is achieved for a window size of 9.}
  \label{tab:resultsWindow}%
\end{table}%

While the impact of the different parameter levels on the overall estimation performance is of interest, we are also interested in how the window size influences the estimation performance. Hence, experiments have been conducted with different window sizes. As the experiments above showed that $LEGO_{ext}$ performed significantly better than $LEGO_{orig}$, only the $LEGO_{ext}$ feature set is used. Again, all experiments are conducted applying 10-fold cross-validation using a linear SVM. The results for UAR, $\kappa$, and $\rho$ are depicted in Figure~\ref{fig:resultsWindow}. Table~\ref{tab:resultsWindow} shows also the relative improvement compared to a window size of three used as baseline for this experiment.

For calculating the window level parameters of all exchanges below the window size, the maximum possible window size of using all exchanges of the respective dialogue are used. Hence, for those exchanges, there is virtually no difference between the parameters on the dialogue and the window level. Of course, this is different for the parameters of exchanges above the window size.

A maximum performance is reached with a window size of 9 for UAR, $\kappa$ and $\rho$ alike. In fact, an UAR of 0.549 represents a relative improvement compared to a window size of 3 by +10.82\%. This clearly shows the potential hidden in these window parameters. If these results are compared to the performance of the original feature set of $LEGO_{orig}$, the performance is even relatively imporved by +15.69\%. This clearly outperforms the currently best know sequential appoach to IQ estimaten applying a Hybrid-HMM~\cite{ultes2014b} by +8.5\%.

It is interesting though that the best window size is nine. We believe that this is system dependent and, in Let's go, related to the minimum number of system-user-exchanges necessary to perform a successful dialogue. Looking at the corpus reveals that a minimum of nine exchanges is needed. 
\vspace{-4pt}

\section{Discussion}
\label{sec:discussion}
When analyzing the results, clearly, the temporal information has a major effect on the Interaction Quality. In fact, the dialogue level parameters contributing most may be interpreted as the satisfaction of the user (represented by IQ) mainly depends on the complete dialogue and not on short-term events. However, putting this long-term information in the context of a shorter more recent period modelled by the window level achieves an even better performance. This increase is even more evident when further adjusting the window size. Hence, it may be concluded that IQ does not purely depend on local effects but those local effects have to be interpreted within the context of the dialogue.
\vspace{-3pt}

\section{Conclusion and Future Work}
\label{sec:conclusion}
In this work, we analyzed the set of temporal parameters used as input for statistical classification approaches to estimate Interaction Quality. We showed that proper modeling of temporal aspects within the feature set may outperform sequential classification approaches like Hidden Markov Models drastically. For some temporal parameters, we introduced both the system and the user view thus extending the feature set. This results in an significant relative increase in UAR by +4.4\%. Furthermore, analysing the temporal features, i.e., features on the dialogue and window level, showed that both levels contribute most to the overall estimation performance (in contrast to the exchange level). Furthermore, we modified the window size further achieving a statistically significant relative improvmnet for IQ estimation by +15.69\% with an UAR of 0.549. The optimal window size of 9 for IQ recognition is attributed to the task complexity. Here, a minimum of 9 exchanges is necessary to successfully complete the task.

More generally, while the Interaction Quality is clearly influenced by local events, the complete course of the dialogue plays a major role. In other words, the user are clearly aware of the complete dialogue and do not ``forget'' events which may occur at the beginning of the dialogue. 

For future work, previous successful experiments improving the estimation performance like applying a Hybrid-HMM or a hierarchical error-correction approach should be investigated using this extended and optimized feature set. Furthermore, repeating the experiments with more data of the \emph{LEGOv2} corpus~\cite{ultes2015} might also give further insight into the problem.


\bibliographystyle{spmpsci}
\bibliography{references}

\end{document}